SH 1.7.02

# Detection of 6 November 1997 Ground Level Event by Milagrito

J.M. Ryan[1]   for Milagro collaboration

[1] *Space Science Center, University of New Hampshire, Durham, NH 03824 USA*

**Abstract**

Solar Energetic Particles from the 6 November 1997 solar flare/CME(coronal mass ejection) with energies exceeding 10 GeV have been detected by Milagrito, a prototype of the Milagro Gamma Ray Observatory. While particle acceleration beyond 1 GeV at the Sun is well established, few data exist for protons or ions beyond 10 GeV.  The Milagro observatory, a ground based water Cherenkov detector designed for observing very high energy gamma ray sources, can also be used to study the Sun.  Milagrito, which operated for approximately one year in 1997/98, was sensitive to solar proton and neutron fluxes above ~5-10 GeV.  Milagrito operated in a scaler mode, which was primarily sensitive to muons, low energy photons, and electrons, and the detector operated in a mode sensitive to showers and high zenith angle muons.  In its scaler mode, Milagrito registered a rate increase coincident with the 6 November 1997 ground level event observed by Climax and other neutron monitors.  A preliminary analysis suggests the presence of >10 GeV particles.

## 1 Introduction:

Particle acceleration beyond 1 GeV at the Sun is well established (Parker, 1957), but its intensity and energy still amazes researchers.  However, few data exist demonstrating acceleration of protons or ions beyond 10 GeV (Chiba et al., 1992; Karpov et al., 1997; Lovell et al., 1998).  The energy upper limit of solar particle acceleration is unknown but is an important parameter, since it relates not only to the nature of the acceleration process, itself not ascertained, but also to the environment at or near the Sun where the acceleration takes place.  The Milagro instrument, a water Cherenkov detector near Los Alamos, NM, is at 2650 m elevation with a geomagnetic cutoff rigidity of ≈3.5 GV.  It is sensitive to solar hadronic cosmic rays from approximately 5 GeV to beyond 1 TeV.  These primary particles are detected via Cherenkov light, produced by the secondary shower particles, as they traverse a large (80 × 60 × 8 m) water-filled pond containing 723 photomultiplier tubes (228 PMTs for the prototype, Milagrito).  This energy range overlaps that of neutron monitors (< 10 GeV) such that Milagro complements the worldwide network of these instruments.  These ground based instruments, in turn, complement spacecraft cosmic ray measurements at lower energies.  This suite of instruments may then be capable of measuring the full energy range of solar hadronic cosmic rays, with the goal of establishing a fundamental upper limit to the efficiency of the particle acceleration by the Sun.

Milagro's baseline mode (air shower telescope mode) of operation measures extensive air showers above 300 GeV from either hadrons or gamma rays.  A description of Milagro's capabilities as a VHE gamma ray observatory is available elsewhere in these proceedings (McCullough, 1999).  Milagro measures not only the rate of these events but also the incident direction of each event, thereby localizing sources.  While performing these measurements, the instrument records the rate of photomultiplier hits (the scaler mode), with an intrinsic energy threshold of about 5 GeV for the progenitor cosmic ray to produce at least one hit.  The scaler mode is similar to that of a neutron monitor, while the telescope mode can significantly reduce background by pointing.  With a proposed fast data acquisition system (DAQ) and modified algorithms for determining incident directions of muons, the energy threshold of Milagro's telescope mode will be reduced to ~10 GeV by detecting the (≈300 kHz) single muons and mini muon showers. For now, this low energy threshold can only be achieved by using Milagro in the scaler mode, which is not capable of localizing sources. A description of the Milagro solar telescope mode was presented earlier (Falcone et al., 1999).

## 2 Solar Milagro/Milagrito Scaler Mode:

In scaler mode, a substantial portion of the rate recorded by Milagro (and Milagrito) is due to muons, and an integral measurement above threshold is performed. These data will provide an excellent high energy complement to the network of neutron monitors, which has been, and continues to be, a major contributor to our understanding of solar energetic particle acceleration and cosmic rays. Monte Carlo events have been used to estimate the effective areas of Milagrito to protons incident on the atmosphere isotropically, at zenith angles ranging from $0^\circ$-$60^\circ$ (see figure 1). The effective area curves for Milagro, which have been plotted for the sake of comparison, are for vertically incident protons. At 10 GeV, Milagro's scaler mode has nearly 100 times the effective area of a neutron monitor, with the effective area rising rapidly with energy, while Milagrito had approximately 4 times the effective area of a neutron monitor at 10 GeV. Pressure, temperature, and other diurnal corrections must be applied to the ground level scaler rate (Hayakawa, 1969). We have begun to determine these correction factors for Milagro/Milagrito, and we find them to be reasonably consistent with past work with muon telescopes (Fowler & Wolfendale, 1961). However, these corrections are less important for transient (i.e. solar) events that rise above background quickly and have short durations.

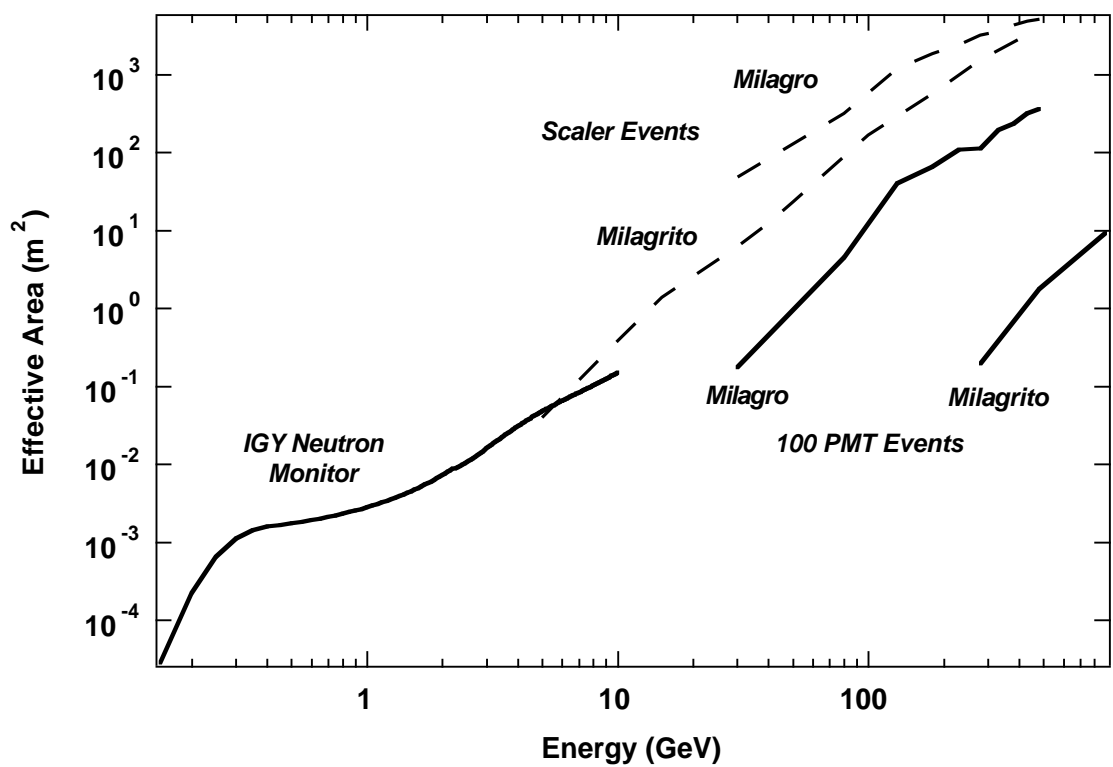

**Figure 1 : Effective area curves for Milagro and Milagrito, with an IGY neutron monitor for comparison.**
(Milagro shower trigger presently requires ≈150 PMTs )

## 3 November 6, 1997 Ground Level Event:

On 6 November 1997 at approximately 12:00 UT, an X-class flare with an associated coronal mass ejection occurred on the Sun. This produced a nearly isotropic ground level event registered by many neutron monitors. A preliminary analysis of neutron monitor data for this proton event yields the spectral index of 5.5 at event maximum, assuming a power law proton spectrum (Smart & Shea, 1998). Climax, located in nearby central Colorado, is the closest of these neutron monitors to Milagro/Milagrito. Milagrito, a prototype version of Milagro with less effective area, registered a scaler rate increase coincident, within error, with that measured by Climax (see Figure 2). If one accounts for the meteorological fluctuations, the event duration and time of maximum intensity, as seen by Milagrito, are also consistent with that of Climax.

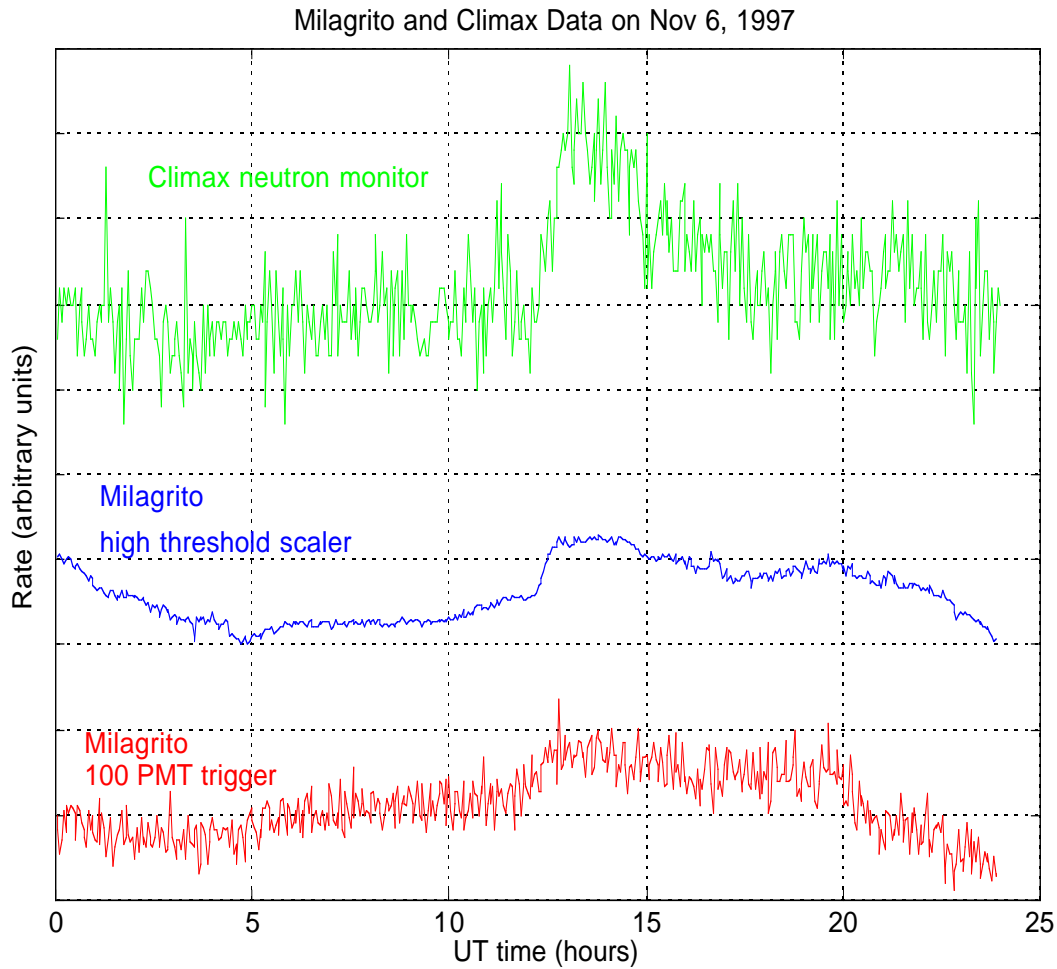

**Figure 2 : Milagrito registered a rate increase coincident with that of Climax during the GLE of Nov. 6, 1997. The y-axis units have been scaled and shifted for each plot to make comparison easier. (Climax data courtesy of C. Lopate, Univ. of Chicago)**

The 100 PMT shower trigger rate also experienced an increase, although the significance is not as great as that in scaler mode. It is not yet clear which of several possible mechanisms initiated the signal in the 100 PMT shower trigger. This increase can be caused by high energy primaries ($> 100$ GeV, see figure 1) or secondary muons arriving from a nearly horizontal direction. If the signal was caused by high energy protons, then it can be compared to the scaler mode rate increase in order to derive a proton spectrum. This is done by integrating a "test sample" power law spectrum of protons multiplied by the effective area for the detector in its two modes. The parameters of the "test sample" are then varied until a good fit to the measured rate increases is achieved. This analysis was done, using an extrapolated form of the 100 PMT mode effective area curve, and the derived power law spectrum has an index of $>7$. This spectrum derived from Milagrito is softer than that of the world wide network of neutron monitors (index $\approx 5.5$), likely indicating a cutoff, or a roll over, somewhere in Milagrito's range of sensitivity. However, it appears as though horizontal secondary muons have contributed to this signal. These muons could still be the result of proton primaries, but the effective area of the detector would be significantly different from that assumed here. Future work will address this issue by recalculating the spectrum by analyzing events caused by horizontally incident secondary muons.

This work is supported in part by the National Science Foundation, U.S. Department of Energy Office of High Energy Physics, U.S. Department of Energy Office of Nuclear Physics, Los Alamos National Laboratory, University of California, Institute of Geophysics and Planetary Physics, the Research Corporation, and the California Space Institute.